\def\cm{cm$^{-1}$}
\documentstyle[prl,aps,multicol]{revtex}
\input{psfig}

\begin{document}
\sloppy
\title{Extremely Small Energy Gap in the Quasi-One-Dimensional \\
Conducting Chain Compound SrNbO$_{3.41}$}
\author{C. A. Kuntscher$,^{1,2}$ S. Schuppler,$^{1}$ P. Haas,$^{2}$
B. Gorshunov,$^{2,*}$ M. Dressel,$^{2}$ M. Grioni,$^{3}$ F. Lichtenberg,$^{4}$ \\
A. Herrnberger,$^{4}$ F. Mayr,$^{4}$ and  J. Mannhart$^{4}$}

\address{
$^1$ Forschungszentrum Karlsruhe, Institut f\"ur
Festk\"orperphysik,
D-76021 Karlsruhe, Germany \\
$^2$ 1. Physikalisches Institut, Universit\"at Stuttgart,
Pfaffenwaldring 57, D-70550 Stuttgart, Germany\\
$^3$ Institut de Physique Appliqu\'ee, Ecole Polytechnique
F\'ed\'erale,
CH-1015 Lausanne, Switzerland \\
$^4$ Institut f\"ur Physik, EKM, Universit\"at Augsburg,
Universit\"atsstr. 1, D-86135 Augsburg, Germany }
\date{Received \today}
\maketitle

\begin{abstract}
Resistivity, optical, and angle-resolved photoemission
experiments reveal unusual one-dimensional electronic properties
of highly anisotropic SrNbO$_{3.41}$. Along the conducting chain
direction we find an extremely small energy gap of only a few meV
at the Fermi level. A discussion in terms of typical 1D
instabilities (Peierls, Mott-Hubbard) shows that neither seems to
provide a satisfactory explanation for the unique properties of
SrNbO$_{3.41}$.
\end{abstract}


\begin{multicols}{2}
\columnseprule 0pt \narrowtext A one-dimensional (1D) interacting
electron system no longer exhibits the quasiparticles known from
Fermi liquid theory and is instead a Tomonaga-Luttinger liquid
(TLL) whose properties also include spin-charge separation and a
power-law behavior of correlation functions \cite{Voit95a}. This
1D theory does not, however, directly apply to {\em real} 1D
materials since some coupling between the 1D entities (such as chains 
or similar) is always present, rendering the systems {\em quasi-}1D
and facilitating 2D and 3D ordering effects. Real 1D materials
are thus conducive to various instabilities masking the predicted
TLL characteristics \cite{Grioni00}. Electron-phonon coupling,
for instance, can lead to a gapped charge-density wave (CDW)
ground state (Peierls transition), and electron-electron
interaction may cause the opening of a Mott-Hubbard (MH) gap at
commensurate band filling.

Recently, a new quasi-1D material was synthesized
\cite{Lichtenberg91} which belongs to the series SrNbO$_{3.5-x}$
of perovskite-related transition-metal oxides, with
characteristics spanning a wide range: Depending on the oxygen
stoichiometry, these niobium oxides exhibit quasi-1D metallic
character ($x$$\approx$0.1) \cite{Kuntscher00} or
ferroelectricity ($x$=0) with very high transition temperature
\cite{Nanamatsu71}. In this Letter we focus on the unusual
electronic properties of the quasi-1D metallic compound
SrNbO$_{3.41}$ which were determined by dc resistivity, optical
spectroscopy, and angle-resolved photoemission (ARPES)\@. Along
the conducting chain direction, an extremely small energy gap of
only a few meV, much smaller than for other quasi-1D compounds,
is observed at the Fermi level. The gap is discussed in terms of
a Peierls and a MH scenario.

\begin{figure}[h]
\centerline{\psfig{file=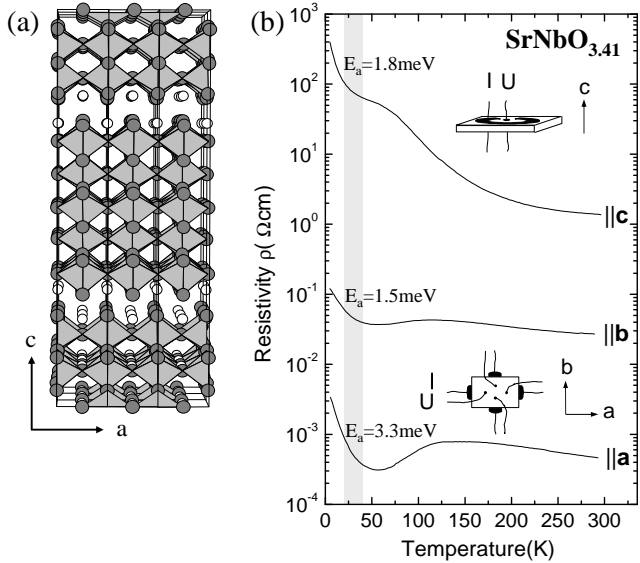,width=85mm,clip=}} 
\caption{(a) Projection along the $b$ axis of the SrNbO$_{3.4}$
crystal structure; Nb atoms are hidden within the NbO$_6$
octahedra (light grey).  White (grey) circles: Sr (O) atoms. (b)
Dc resistivity $\rho$ versus temperature of SrNbO$_{3.41}$ along
the three crystal axes, with the resp.\ measurement geometry. In
the range 20-40 K (grey bar) one finds an activated behavior
$\rho\propto\exp\{E_a/k_BT\}$ with $E_a$ as indicated.}
   \label{fig:dc}
\end{figure}

SrNbO$_{3.41}$ single crystals of typical size
3$\times$2$\times$0.2 {\rm mm}$^3$ were grown
\cite{Lichtenberg91,Lichtenberg01} by the floating zone melting
technique. The precise oxygen content was determined
thermogravimetrically. The analysis of the room temperature (RT)
x-ray powder diffraction pattern indicates single-phase
composition with lattice constants $a=3.99$~\AA, $b=5.67$~\AA,
and $c=32.46$~\AA, in good agreement with earlier work
\cite{Abrahams98}. The basic structural building blocks are
NbO$_6$ octahedra [see Fig.\ \ref{fig:dc}(a)] grouped into slabs
which extend parallel to the ($a,b$) plane and for SrNbO$_{3.4}$
are five octahedra wide along $c$. Only along the $a$ axis the
octahedra are connected continuously via their apical oxygen
atoms, forming 1D chains.

The dc resistivity of SrNbO$_{3.41}$ along the three axes was
measured using a four-point configuration \cite{Lichtenberg01}.
Polarized reflection measurements along the $a$ and $b$ axis were
performed for frequencies 6-34$\,$000~\cm\ by utilizing various
spectrometers; for details see Ref.\ \cite{Kuntscher02}. To
accomplish the Kramers-Kronig analysis, we used a Hagen-Rubens
extrapolation for $\omega$$\rightarrow$0 and a power law
($\omega^{-4}$) at high frequencies. High-resolution
($\Delta$$E$=15~meV) ARPES data were recorded with a Scienta
ESCA-300 analyzer using the He I$\alpha$ line of a discharge lamp.
The samples were cleaved at RT and a base pressure of
1 $\times$ 10$^{-10}$ mbar to expose an (001) surface, and cooled
down to 25 K. By Laue diffraction the crystals were oriented to
$\pm$1$^{\circ}$. The angular resolution was set to
$\pm$0.5$^{\circ }$, and the Fermi level $E_F$ was determined to
1 meV accuracy from the Fermi cutoff of a freshly evaporated Au
film recorded immediately afterwards at the same experimental
conditions.

Fig.~\ref{fig:dc}(b) shows the dc resistivity $\rho$ versus
temperature $T$ of SrNbO$_{3.41}$ along the three axes. It is
highly anisotropic with very low, metal-like values along the $a$
direction ($\rho_a$(300K)=4.6$\times$10$^{-4}$ $\Omega$cm) and a
RT anisotropy of $\rho_a$:$\rho_b$:$\rho_c$=1:10$^2$:10$^4$. The
variation of $\rho$ with temperature is smaller within the
($a,b$) plane than along $c$. In all directions a broad maximum
or shoulder is seen between 200 and 50~K, and below 50 K we find
an increase of $\rho$, which in the range 20-40 K can be
described by an activated behavior $\rho$$\propto$$\exp\{E_a/k_BT\}$ 
with activation energies $E_a$=2-3~meV. In the range 60-130 K $\rho_a$ 
(along the chains) has a metallic $T$ dependence, whereas above 130 \nolinebreak K 
it slightly decreases with increasing temperature 
(thermally activated hopping between metallic strands may explain this). 
A similar temperature dependence is observed for $\rho_b$, however, 
not as clearly developed. There are no signs of metallic behavior along
$c$.

\begin{figure}
\centerline{\psfig{file=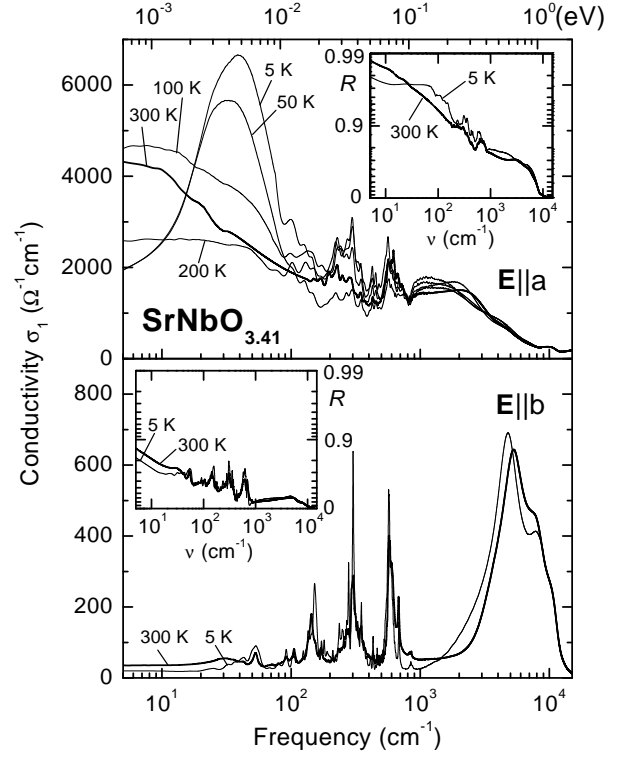,width=80mm,clip=}}
\caption{Frequency dependent conductivity of SrNbO$_{3.41}$ for
different temperatures and the polarization {\bf E} set parallel
to the $a$ and $b$ axes (insets: reflectivity $R$ used for
Kramers-Kronig analysis). For {\bf E}$||$$a$ the low-temperature
peak around 40 \cm\ indicates excitations across a gap
2$\Delta$$\approx$5 meV. }
   \label{fig:optics}
\end{figure}

The anisotropy in the electrical properties of SrNbO$_{3.41}$ is
also clearly seen in the optical response (Fig.~\ref{fig:optics}).
The reflectivity $R$ displayed in the insets of
Fig.~\ref{fig:optics} shows a sharp plasma edge for the
polarization {\bf E}$\parallel$$a$ with high values
($R$$\approx$1) at low frequencies, indicating metallic
conductivity. In contrast, for {\bf E}$\parallel$$b$ the material
is insulating and we mainly see phonon contributions between 40
and 1000~\cm\ showing significant changes with temperature which
will be discussed elsewhere \cite{Kuntscher02}. For both
polarization directions the low frequency ($\omega$$\rightarrow$0)
conductivity $\sigma_1$ agrees, in general, well with the
measured dc data (Fig.~\ref{fig:dc}), reproducing its rather
complicated temperature behavior. At RT the {\bf E}$\parallel$$a$
conductivity contains a Drude-like contribution at low frequencies
followed by several phonon lines between 100 and 1000~\cm\ and a
broad mid-infrared band around 1500~\cm. Upon reducing the
temperature, $\sigma_1$ exhibits substantial changes at
frequencies lower than 100~\cm, leading to the appearance of a
peak around 40~\cm, which is already quite strong at 50~K\@. At 5K, 
this peak has shifted slightly to higher frequencies and has
grown even stronger. We ascribe this feature to single-particle
excitations across an energy gap in the electronic density of
states with 2$\Delta$(5 K)$\approx$5 meV.

The momentum resolved electronic properties probed by ARPES
reveal a quasi-1D band structure for
Sr$_{0.9}$La$_{0.1}$NbO$_{3.39}$ \cite{Kuntscher00} as well as
for SrNbO$_{3.41}$ \cite{Kuntscher02}: A dispersing band is found
only along {\it one} high-symmetry line, $\bar\Gamma$-$\bar X$,
parallel to the chain direction $a$, and seems to cross $E_F$; a
second band near $E_F$ does not show discernible dispersion.
Neither band disperses along the transverse direction
$\bar\Gamma$-$\bar Y$. The Fermi surface (FS) is 1D
\cite{Kuntscher00}, with nesting vector 2$k_F$ of about 1/3 of
the Brillouin zone.

Further insight is gained by band structure calculations in the 
local density approximation
\cite{Kuntscher00,Winter00} which demonstrate that among the
three inequivalent Nb sites per slab a predominant
\cite{Winter00} contribution to the density of occupied states at
$E_F$ comes from the {\it central} Nb site; the contributions of
the other two Nb sites rapidly fall off with distance away from
the center. Distortion is small only for the NbO$_6$ octahedra
based on the central Nb sites, and one can thus clearly relate
the quasi-1D character of the dispersing band to these chains of
almost ``perfect'' octahedra in the middle of the slabs. The
importance of the central chains is also demonstrated by
comparing ARPES results for SrNbO$_{3.41}$ with those for
Sr$_{0.8}$La$_{0.2}$NbO$_{3.50}$ \cite{Kuntscher02}, where the
central chains are missing since there the slabs are only {\it
four} octahedra wide: for the latter the 1D character of the band
structure has almost vanished.

\begin{figure}
\centerline{\psfig{file=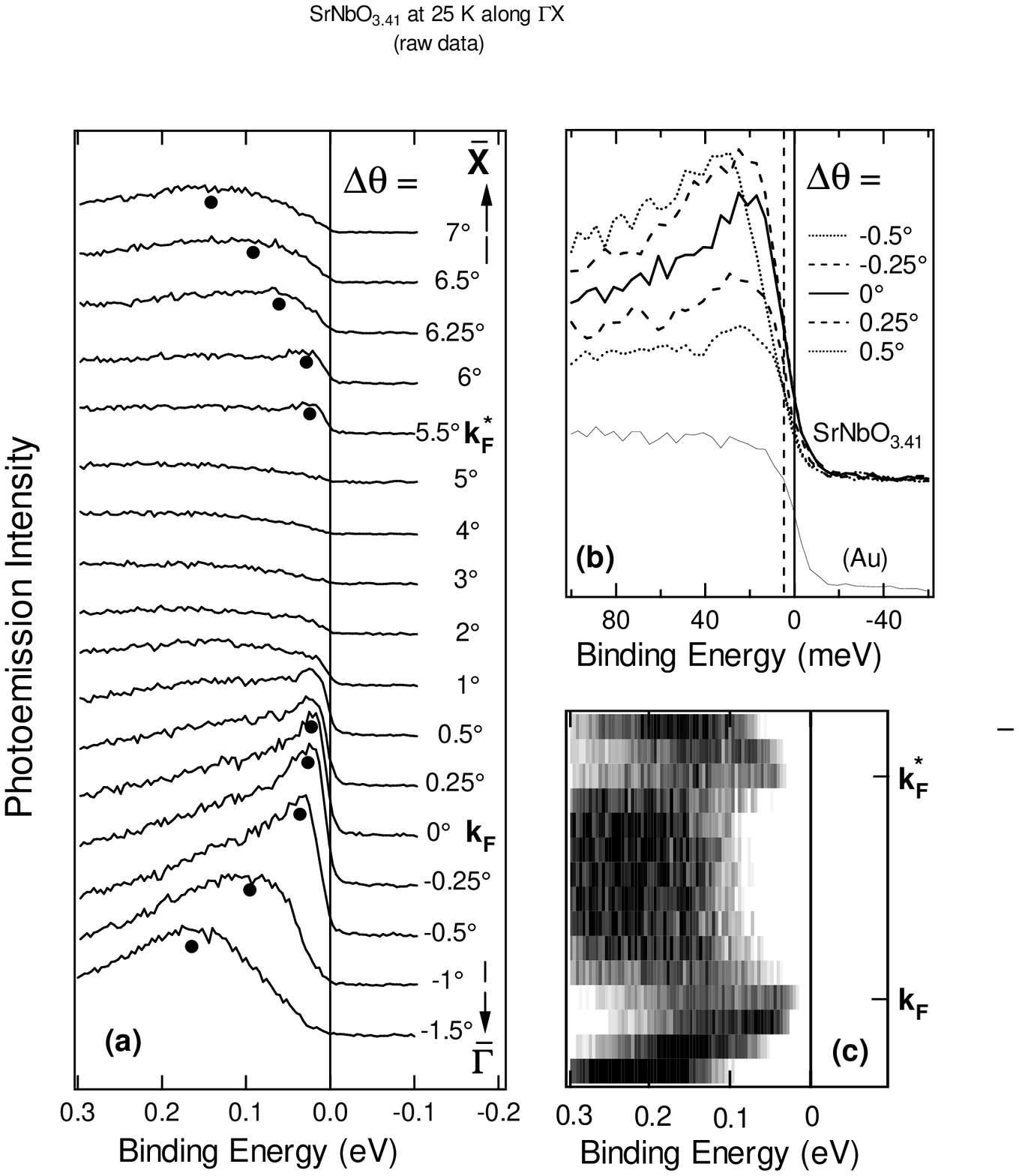,width=80mm,clip=}}
\caption{ARPES spectra of SrNbO$_{3.41}$ along $\bar\Gamma$-$\bar
X$ near $k_F$ at $T$=25 K.
  The angle $\Delta$$\theta$ is the emission angle with respect to $k_F$
  ($\Delta$$\theta$$\equiv$0$^{\circ}$).
  (a) Spectra between the two ``crossing points'' ($k_F$,$k^*_F$) of the
  strongly dispersing band; peak positions are marked by dots.
  (b) Spectra around $k_F$ in 0.25$^{\circ}$ steps. A gap $\Delta$$\approx$5 meV
  is seen when compared to the Fermi cutoff of a gold film (thin solid line).
  (c) Intensity plot of spectra, taken between $k_F$ and $k^*_F$ (step size
  0.5$^{\circ}$) and normalized to constant total intensity; a weak, dispersing
  shadow band can be identified.
}
      \label{fig:ARPES}
\end{figure}

The quasi-1D FS with almost complete nesting makes SrNbO$_{3.41}$
prone to instabilities opening a gap at $E_F$, and we thus
concentrate on the region around $k_F$ in order to very
sensitively detect such signatures. Fig.~\ref{fig:ARPES} (a)
shows the corresponding high-resolution ARPES spectra along
$\bar\Gamma$-$\bar X$ near $E_F$. At
$\Delta$$\theta=-1.5^{\circ}$ the peak associated with the
strongly dispersing band is located at $\approx$160 meV binding
energy and is rather broad, but sharpens up as it approaches
$E_F$ and appears to cross it. We define $k_F$
($\Delta$$\theta$$\equiv$0$^{\circ}$) as the emission direction
for which the leading edge midpoint of the spectrum is closest to
$E_F$. Due to a 2$\times$1 superstructure \cite{Kuntscher00} the
peak reappears already at $k_F^*$=5.5$^{\circ}$ and disperses
towards higher binding energies for increasing $\Delta\theta$.
Fig.~\ref{fig:ARPES} (b) depicts the ARPES spectra around $k_F$
for a narrow angular step size of 0.25$^{\circ}$ together with
the Fermi cutoff of a Au film. Starting from $-0.5^{\circ}$ and
going to higher $\Delta$$\theta$, the leading edge midpoint
approaches $E_F$ but {\it does not reach it} for any of the
spectra. At $k_F$, the spectral weight at the Fermi level is
significantly suppressed (by $\approx$40\%) compared to the
leading edge midpoint, and we find a gap between the leading edge
midpoint and $E_F$ of $\Delta$(25 K)$\approx$4 meV. Although this
``leading edge'' method is known to underestimate the gap size
somewhat for a peak-like structure a fit (not shown) of the $k_F$
spectrum, similar to Ref.\ \cite{Armitage01} and accounting for
the peak structure, gives only a slightly larger result,
$\Delta$(25 K)$\approx$(5$\pm$2) meV, corroborating the leading
edge result. Possible extrinsic effects such as charging or too
coarse an angular step size can also be excluded as a source of
the gap \cite{comment00}. The gap is thus clearly established and
with a size of about 5 meV is, in fact, the smallest one found in
ARPES for any quasi-1D compound.

The intensity plot of high-resolution ARPES spectra in
0.5$^{\circ}$ steps and normalized for clarity to constant total
intensity is shown in Fig.~\ref{fig:ARPES} (c). It indicates
additional states between $k_F$ and $k_F^*$, where the strongly
dispersing band is unoccupied: beyond $k_F$ a ``shadow band'' is
visible which disperses away from the gap edge and exhibits a
symmetry consistent with the 2$\times$1 superstructure. Its small
intensity is compatible with the small gap size \cite{Voit00}. As
shadow bands are rather weak in general a clear identification in
experiment has been possible only for very few quasi-1D materials
\cite{Voit00,Vescoli00}.

The different experiments on SrNbO$_{3.41}$ all yield clear
evidence for an energy gap at $E_F$ along the conducting chains:
a rise in the dc resisitivity, a strong peak in the optical
conductivity, as well as a significant suppression of spectral
weight in ARPES\@. All results lead to characteristic energies of
a few meV -- agreeing well enough to demonstrate that the gap does 
exist and suggesting a common origin. Some spread is present, as 
expected, due to the different
nature of the excitation for the various techniques (the optical
gap, e.\ g., is somewhat smaller, possibly due to excitonic
effects).

Most surprising is the gap's {\em extreme smallness}, making it
unique among quasi-1D materials; the unusual properties of the
gap, however, turn out to defy easy explanation: (i) A lattice
superstructure with sufficiently small displacement amplitudes
would of course be able to reproduce the gap and its small size.
It can{\em not}, however, plausibly explain why the gap should be
situated right at $E_F$, considering the large width ($\gtrsim$2
eV \cite{Kuntscher00,Winter00}) of the quasi-1D band. The
instabilities of quasi-1D systems, on the other hand, by their
very nature cause a gap located at $E_F$ and thus are more likely
candidates for an explanation of the gap properties observed for
SrNbO$_{3.41}$ -- such as (ii) a MH-type insulating state with a
gap at $E_F$ due to electron-electron interaction; for this, the
filling level of the quasi-1D band would have to be commensurate,
1/3. This 1/3 filling, however, would require long-range
interaction and a TLL parameter $K_{\rho}<$1/3 for a MH gap to
open \cite{Schulz94}, and the very strong correlation implied by
the latter appears implausible for SrNbO$_{3.41}$: correlation
effects are generally small for $4d$ systems like niobium oxides.
The possibility of a MH gap is thus unlikely yet cannot be fully 
excluded based on our data as theoretical results for 1/3 filling 
are still scarce. (iii) A Peierls instability caused by
electron-phonon coupling appears possible, too, leading to a CDW
state for sufficiently low temperatures with a gap at $E_F$
\cite{Peierls55,transition}\@. The gap size is related to the
mean-field (MF) transition temperature $T_P^{MF}$, and in the
weak-coupling limit 2$\Delta$(0) = 3.52$k_BT_P^{MF}$; from the
ARPES gap 2$\Delta$(25 K)$\approx$10 meV one can then estimate
$T_P^{MF}$$\approx$40 K \cite{meanfield}\@. {\em Only below} this
temperature the gap (or its precursor, the fluctuation-induced
pseudogap) can exist; in this sense $T_P^{MF}$ strictly limits
the temperature range where any gap-related features can possibly
be observed. Yet for SrNbO$_{3.41}$ one clearly sees the upturn
in $\rho_a(T)$ (Fig.~\ref{fig:dc}) to start at $\approx$55 K, and
the low-energy peak for {\bf E}$\|$$a$ in optics
(Fig.~\ref{fig:optics}) is almost fully developed already at 50
K, suggesting an onset well above that temperature. A simple
Peierls picture therefore is also insufficient to provide an
explanation for the observed small gap \cite{ferro}\@. 
Inconsistencies between the experimental results and theorical
models have been found for other quasi-1D systems as well, like the paradigmatic 
(NbSe$_4$)$_3$I, (TaSe$_4$)$_2$I, and K$_{0.3}$MoO$_3$. 
The latter two are CDW systems yet show clear discrepancies to the 
Peierls picture (e. g., ARPES line shape and Fermi cutoff above T$_P$), 
which were attributed to correlation effects \cite{Grioni00,Perfetti01}\@.
In contrast, quasi-1D SrNbO$_{3.41}$ has no obvious Peierls transition, 
little tendency to correlation,  
and the relevant energy scale is drastically smaller. It seems to 
be a system with characteristics distinctly its own, thus expanding 
the parameter space where real 1D systems remain to be fully 
understood.

In summary, we consistently observe in ARPES, dc resistivity, and
optics that SrNbO$_{3.41}$ is a compound with 1D electronic
characteristics around $E_F$; as expected for such a system,
instabilities lead to an energy gap at $E_F$\@. Most surprising is
its small gap size: only a few meV by all techniques and thus
much smaller than for other quasi-1D compounds. Further analysis
shows the experimental findings to appear inconsistent with
validity ranges for both the Peierls and Mott-Hubbard picture.
These aspects promise more general implications for the
understanding of {\em real} 1D systems.

We thank B. G\"otz,  G. Hammerl, A. Loidl, L. Perfetti, C. Rojas,
C. Schneider, R. Schulz, I. Vobornik, J. Voit, and H. Winter for
valuable help and fruitful discussions. This work was supported
by the DAAD, the BMBF (project No. 13N6918/1), and the DFG.

\end{multicols}

\begin{thebibliography}{99}
\bibitem{meanfield}
A BCS-type approximation for $\Delta$($T$) was used for the
estimate.
\bibitem{ferro} The structural affinity of SrNbO$_{3.41}$ to
ferroelectric SrNbO$_{3.50}$ may be of importance: the
central chains responsible for the 1D electronic character are
coupled to a network of polarizable NbO$_6$ octahedra
structurally equivalent to those giving rise to the high
dielectric constant in the ferroelectric, and a certain influence
on the electronic properties of the chains is conceivable.
\end{thebibliography}
\end{document}